\documentclass[usenatbib, twocolumn, a4paper,nofootinbib]{aastex63}

\usepackage[utf8]{inputenc}  
\usepackage[T1]{fontenc}
\usepackage{graphicx,psfrag}
\usepackage{mathrsfs}
\usepackage{amsmath,amsfonts,amssymb,pifont,gensymb}
\usepackage{multirow,enumerate}
\usepackage{comment,hyperref}
\usepackage{color}
\usepackage{acronym}
\usepackage{xspace}
\usepackage[normalem]{ulem}
\usepackage{mathtools}
\usepackage{subfigure}
\usepackage{makecell}
\usepackage{changepage}
\usepackage{appendix}
\usepackage{lipsum}
\usepackage{natbib}

\begin{document}

\title{On the identification of individual gravitational wave image types of a lensed system using higher-order modes} %

\author{Justin Janquart$^{*}$}
\email{$^{*}$j.janquart@uu.nl}
\affiliation{Nikhef – National Institute for Subatomic Physics, Science Park, 1098 XG Amsterdam, The Netherlands }
\affiliation{Institute for Gravitational and Subatomic Physics (GRASP), Department of Physics, Utrecht University, Princetonplein 1, 3584 CC Utrecht, The Netherlands}

\author{Eungwang Seo}
\affiliation{Department of Physics, The Chinese University of Hong Kong, Shatin, N.T., Hong Kong}

\author{Otto A. Hannuksela}
\affiliation{Nikhef – National Institute for Subatomic Physics, Science Park, 1098 XG Amsterdam, The Netherlands }
\affiliation{Institute for Gravitational and Subatomic Physics (GRASP), Department of Physics, Utrecht University, Princetonplein 1, 3584 CC Utrecht, The Netherlands}
\affiliation{Department of Physics, The Chinese University of Hong Kong, Shatin, N.T., Hong Kong}

\author{Tjonnie G. F. Li}
\affiliation{Department of Physics, The Chinese University of Hong Kong, Shatin, N.T., Hong Kong}
\affiliation{Institute for Theoretical Physics, KU Leuven, Celestijnenlaan 200D, B-3001 Leuven, Belgium}
\affiliation{Department of Electrical Engineering (ESAT), KU Leuven,
Kasteelpark Arenberg 10, B-3001 Leuven, Belgium}

\author{Chris Van Den Broeck}
\affiliation{Nikhef – National Institute for Subatomic Physics, Science Park, 1098 XG Amsterdam, The Netherlands }
\affiliation{Institute for Gravitational and Subatomic Physics (GRASP), Department of Physics, Utrecht University, Princetonplein 1, 3584 CC Utrecht, The Netherlands}

\begin{abstract}
\noindent
Similarly to light, gravitational waves can be gravitationally lensed as they propagate near massive astrophysical objects such as galaxies, stars, or black holes. In recent years, forecasts have suggested a reasonable chance of strong gravitational-wave lensing detections with the LIGO-Virgo-KAGRA detector network at design sensitivity. As a consequence, methods to analyse lensed detections have seen rapid development. However, the impact of higher-order modes on the lensing analyses is still under investigation. In this work, we show that the presence of higher-order modes enables the identification of the individual images types for the observed gravitational wave events when two lensed images are detected, which would lead to unambiguous confirmation of lensing. In addition, we show that higher-order mode content can be analyzed more accurately with strongly lensed gravitational wave events. 
\end{abstract}

\section{Introduction}
\label{sec:Intro}
Similar to an electromagnetic wave, a gravitational wave (GW) can be deflected by a massive object along its path. This object is called a lens, and depending on its characteristics, it will have a different effect on the GW. For massive lenses, such as galaxies~\cite{Dai:2016igl, Ng:2017yiu, Li:2018prc, Oguri:2018muv} or galaxy clusters~\cite{Smith:2017mqu,Smith:2018gle,Smith:2019dis,Robertson:2020mfh,Ryczanowski:2020mlt}, one can observe strong lensing, where several images of the GW are produced. These images will appear in the interferometers as repeated events with the same frequency evolution. However, the images have a different apparent luminosity distance (linked by a relative magnification), time of coalescence (linked by a time delay, ranging from seconds to months depending on the lens properties), and an overall phase shift (determined by the so-called Morse factor) due to being focused by the lens in slightly different ways~\cite{Wang:1996as, Haris:2018vmn}. The Morse factor is a discrete parameter with three possible values: 0, 0.5, and 1, corresponding to so-called type I, type II and type III images respectively~\cite{Dai:2017huk}. 

Strong lensing is predicted to be observable with a rate of $\mathcal{O}(1)$ per year in  a network of 2G detectors at design sensitivity~\cite{Ng:2017yiu, Li:2018prc, Oguri:2018muv, Xu:2021bfn, Wierda:2021upe}. For example, Ref.~\cite{Wierda:2021upe} predicts $1.7^{+0.9}_{-0.6} \, \rm{yr}^{-1}$ for a LIGO-Hanford, LIGO-Livingston~\cite{TheLIGOScientific:2014jea}, Virgo~\cite{TheVirgo:2014hva} and KAGRA~\cite{Somiya:2011np, Aso:2013eba, Akutsu:2018axf, Akutsu:2020his} 
network.  
Prompted by this, search techniques for strong lensing have been developed over the last years~\cite{Haris:2018vmn, Dai:2020tpj, Liu:2020par, Lo:2021nae, Janquart:2021qov}, and several searches for lensing signatures in the LIGO-Virgo data have been conducted~\cite{Hannuksela:2019kle, Dai:2020tpj, Liu:2020par, LIGOScientific:2021izm}. 
The science that would be enabled by the observation of lensed events spans the domains of fundamental physics, astrophysics, and cosmology~\cite{Sereno:2011ty,Baker:2016reh,Fan:2016swi,Liao:2017ioi,Lai:2018rto,Cao:2019kgn,Li:2019rns,Mukherjee:2019wfw,Mukherjee:2019wcg,Goyal:2020bkm,Diego:2019rzc,Hannuksela:2020xor,Oguri:2020ldf,Cremonese:2021puh,Finke:2021znb}. 

In recent years there have also been efforts to improve the quality of the waveform models used in the analysis of GWs, among other things by adding higher-order modes (HOMs)~\cite{KumarMehta:2019izs, Pratten:2020ceb, Garcia-Quiros:2020qpx}, as their non-inclusion could lead to a loss in sensitivity, or biases in the estimated parameters~\cite{CalderonBustillo:2015lrt, Varma:2016dnf, LIGOScientific:2016ebw}. In addition to improvement in parameter estimation accuracy~\cite{VanDenBroeck:2006ar, Arun:2007hu, Arun:2008xf, London:2017bcn, PhysRevD.99.124005}, HOMs could also enable improved test of general relativity performed with GWs~\cite{PhysRevD.98.024019, Lasky:2016knh, Talbot:2018sgr, Dhanpal:2018ufk, Islam:2019dmk}.

The impact of HOMs on lensing was recently highlighted in Ref.~\cite{Ezquiaga:2020gdt}. In addition to a degeneracy with the antenna pattern functions, binary black hole signals (BBHs) without HOM taken into account have a degeneracy between the Morse phase and the phase of coalescence, so that the image type cannot be determined. However, this degeneracy can be lifted when HOMs are present, leading to the possibility of measuring the Morse factor. 
In \cite{Wang:2021}, the authors studied the possibility of identifying single type II images for current and future detectors. In  \cite{Aditya:2021}, it was shown that not including the Morse factor in analyzing a type II image with significant HOMs leads to biases in the inferred parameters; moreover, given sufficient power in the HOMs, the image type can be identified. However, all these analyses focus on single images and require strong HOM contributions. A first demonstration of the possibility to identify image types based on two images was performed in \cite{Lo:2021nae}, using a single example. 

Here we go considerably further, by exploring a range of HOM contributions for the two images, and investigating more generally when the image identification is possible. We will also demonstrate how ignoring HOMs in the template waveforms could lead us to miss the detection of a lensed event. Moreover, we will investigate whether the observation of a lensed image pair would help in studying the HOMs present in BBH signals.

\section{Methodology}
\label{sec:Methodology}
Our first objective is to understand under what circumstances HOMs will enable us to identify the types of images present in an observed lensed image pair. To explore the effect of different HOM contributions, we tune the HOM signal-to-noise ratio (SNR) by varying the mass ratio $q = m_2/m_1$ (with $m_1, m_2$ the component masses), the inclination $\iota$, and the luminosity distance $D_L$ of the events.  For the first image, $D_L$ is adapted so that the network SNR~\cite{Sathyaprakash:2009xs} for the event is always $12$. For the mass ratio, the three values considered are $0.1, 0.3, 0.5$, while for the inclinations we choose the values to be $20\degree, 45\degree$, and $70\degree$. When considering lensing, we also need to specify the image types, as well as the relative magnification $\mu_{rel}$ and time delay $\Delta t$ between the two images. We consider three types of lensed systems: type I--type II, type I--type I, and type II--type II, where a type I image has a Morse factor $n = 0$, and a type II image has $n = 0.5$. One could also have type III images, with $n = 1$. However, these are expected to be rare, as they would require lenses with very shallow central profiles~\cite{Collett:2015tma, 2013ApJ...773..146D, Collett:2017ksf}, and are therefore not considered here. Throughout this work, the time delay between the two images is arbitrarily fixed at $11\,\rm{hr}$ while the relative magnification is such that the SNR of the second image has a specific value; we consider values of 12 or 25 for the second image as these represent, respectively, a typical and a loud event based on current LVK observing runs~\cite{LIGOScientific:2021usb}. The other parameters are set to arbitrary values, and kept identical for all the simulated events unless stated otherwise. 

For each event, we inject the GW with the \textsc{IMRPhenomXHM} waveform model~\cite{Garcia-Quiros:2020qpx} in a network of interferometers made of the two LIGO detectors and the Virgo detector at design sensitivity~\cite{aLIGOdesign, TheVirgo:2014hva} assuming Gaussian, stationary noise, and we perform the analyses with \textsc{IMRPhenomXHM} and \textsc{IMRPhenomD}~\cite{Khan2016} as template waveforms. To this effect, we use the joint parameter estimation framework laid out in Ref.~\cite{Janquart:2021qov} for analyzing multiple lensed images. The main idea behind this framework is to use the posterior from one image as the prior for another image, which together with the use of a lookup table leads to a significant speed-up in the analysis.
ss
The priors used in our analyses and the general setup are the same as in Sec. 6 of ~\cite{Janquart:2021qov}. That is, the priors for the lensing parameters are uniform for the relative magnification and time delay, and discrete uniform for the (difference in) Morse factor. Furthermore, we choose a uniform prior for the chirp mass, the mass ratio, coalescence time, the cosine of the inclination angle, the polarization angle, and the coalescence phase. The prior for the sky position is such that we have a uniform distribution for the location on a sphere, and the luminosity distance prior is uniform in  comoving volume.

\section{Results}
\label{sec:Results}

Here we first look at the possibility of identifying the individual image types for an observed pair of lensed images. We investigate how our ability to do so evolves with the HOM content of the image pair, and contrast this with the scenario where only one image is detected. We also look at the impact of analyzing an event pair with HOMs using a waveform that does not include them. Finally, we investigate whether our ability to discern the HOM content (and not only the image type) improves when we analyse two images jointly.

\subsection{Type I--type II systems}
\begin{figure}
    \centering
    \includegraphics[keepaspectratio, width=0.5\textwidth]{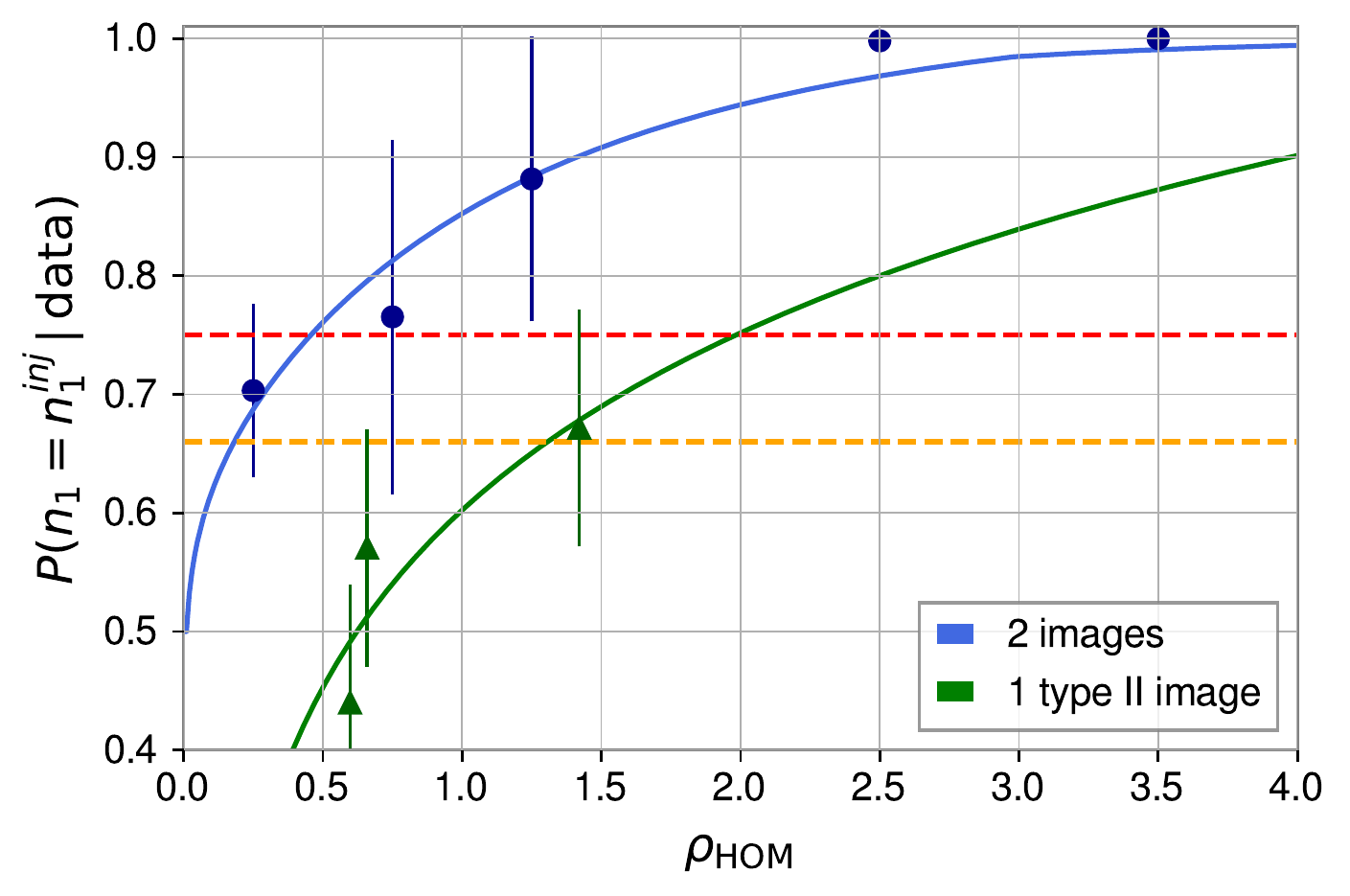}
    \caption{Comparison between the posterior probability values for the recovery of the Morse factor, for systems made of a type I and a type II image (in blue), and for a single type II image (in green) as a function of the total SNR in the HOMs. For the lensed system, the first image corresponding to the type I image has a fixed SNR of 12, while the second image has an SNR of 12 or 25. We change the HOM content of the images by using different combinations of mass ratios and inclinations, and show medians and 90\% intervals for the distribution of probabilities. For the single image systems, the SNR is fixed at 25 and we again change the HOM content by varying the mass ratio and the inclination. The image type identification is made at lower total HOM content when two images are observed than when only one type II image is observed. The recovery for the type II image is the same as for the type I image, as the difference between the Morse factor is always unequivocally recovered.}
    \label{fig:2imgs_Vs_1img}
\end{figure}

First we consider a system of type I and type II images and investigate our ability to recover the image types; we contrast this with the case of a single type II image. Note that when performing a joint analysis on two images, the \emph{difference} in the Morse phase can always be unambiguously determined for the systems considered here (with $\Delta n = 0.5$ for the image configuration at hand). From this information one can infer straight away that the first image is not a type III image. Next, as a  heuristic criterion to determine that an image type is correctly recovered in the two-image case, we choose that the posterior probability $P(n_1 = n_{1}^{inj} | \rm{data}) \geq 0.75$, where $n_1^{inj}$ is the injected value of the Morse factor of the first image. Indeed, when we have no information at all about the image type, we expect a probability of 0.5 for both image types. The value of 0.75 corresponds to half of the probability of the disfavored image type going to the correct one. We find this to be the case, on average, once the HOM SNR (defined as the quadrature sum of the SNRs over individual modes and over the two images) satisfies $\rho_{\rm{HOM}} \gtrsim 0.5$. On the other hand, for a single image we cannot immediately discard the type III image scenario, so that there are now three image types to consider. When we have no information about image types, each of these come with a probability of 0.33. Here we choose $P(n_1 = n_{1}^{inj} | \rm{data}) \geq 0.66$ as the (again  heuristic) criterion to determine that the event type is correctly identified in the case of single images; as before, this corresponds to half of the probability of the disfavored image types going to the correct one. This threshold is crossed when $\rho_{\rm{HOM}} \gtrsim 1.3$. Consequently, for a lensed event pair, the identification of the image types can be done at a weaker HOM contribution than for a single type II image. A comparison of the way $P(n_{1} = n_{1}^{inj} | \rm{data})$ evolves with the HOM SNR can be seen in Fig.~\ref{fig:2imgs_Vs_1img}: The decision threshold is crossed for a lower $\rho_{\rm{HOM}}$ when two images are observed. 

We note that the unequivocal recovery of the image types for an event pair would constitute smoking-gun evidence for lensing, as no other ``standard'' effect could reproduce similar results \cite{Ezquiaga:2020gdt, Wang:2021}.

\subsection{Type I--type I and type II--type II systems}

Let us now consider other types of systems, namely type I--type I and type II--type II. 

The ability to identify the image types for a given pair depends on the types of images present. As in the previous case, when two type II images are detected, we find that the image types can be identified at a lower total HOM SNR than for the observation of a single type II image.

On the other hand, we cannot identify the image types unequivocally for type I--type I systems, regardless of the HOM SNR. When the HOM SNR is high enough, it is possible to exclude the presence of a type II image, and we can say that the two images are of the same type. However, it will be difficult to distinguish type I-type I from type III-type III systems on the basis of GW data alone. That said, type III imaging is expected to be rare when considering a galaxy lens \cite{Collett:2015tma, 2013ApJ...773..146D, Collett:2017ksf}, so in that sense the interpretation of two type-I images will be preferred. On the other hand, the situation regarding type III is less clear when galaxy cluster lenses are considered.

These observations show that we will require at least one type II image to determine the image types based on GW data alone.

\subsection{Note about the use of templates without HOMs when analyzing a system with HOMs}
Without HOMs, the coalescence phase and the Morse phase are degenerate; hence image type identification is not possible when using template waveforms without HOMs~\cite{Ezquiaga:2020gdt}. In addition, the non-inclusion of the HOM in the analysis of events containing significant HOMs can lead to biases in some parameters such as the polarization angle, the phase, and the distance~\cite{CalderonBustillo:2015lrt, Varma:2016dnf}. Since this bias will change depending on the antenna response of the detector, the two images making up the lensed system are biased differently. And indeed, for a type I--type II system, when the HOM content is strong (e.g. $\rho_{\rm{HOM}} = 3.5$), our framework is not able to detect lensing any longer when the analysis is done with \textsc{IMRPhenomD}.

\subsection{Improved probing of HOMs with lensing}
Finally, we compare parameter estimation results for a type I--type II image system with those for a single unlensed image, both having the same total SNR (with a value of $16.97$ for $\sqrt{\sum_{i = 1,2} \rm{SNR}_{i}^{2}}$, where $i$ runs over the images) and $\rho_{\rm{{HOM}}} / \rho_{\rm{tot}} = 0.14$. We use the same BBH parameters for the different types of systems, except for the polarization angle and the (apparent) luminosity distance. In that sense the total HOM content is the same in both scenarios, enabling us to probe whether observing a lensed pair of events leads to  better inference on the HOMs. 

 \begin{figure}
     \centering
     \includegraphics[keepaspectratio, width=0.49\textwidth]{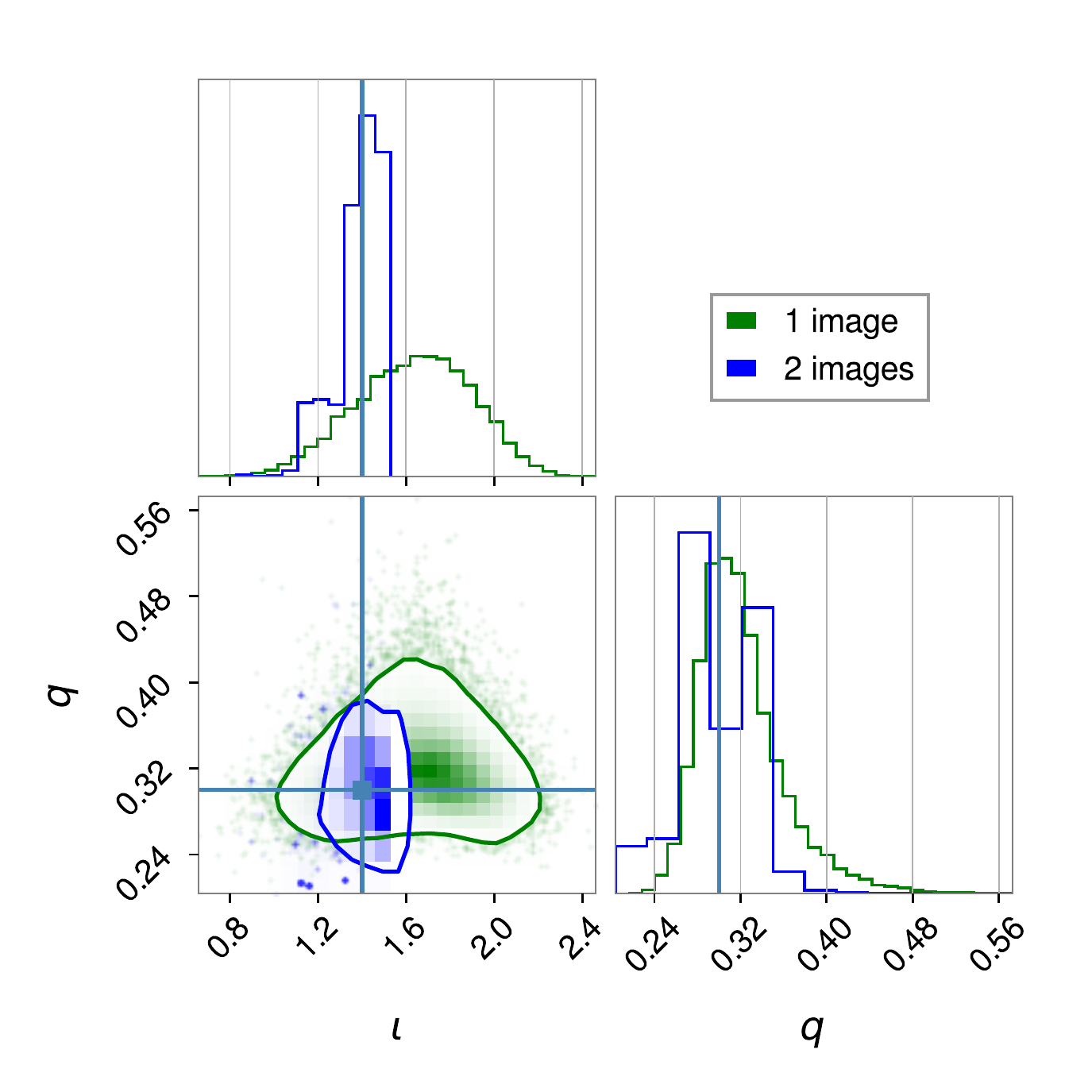}
     \caption{The posterior distribution for the inclination ($\iota$) and the mass ratio ($q$) for an unlensed event (in green) and a lensed image pair (in blue). The events have the same (total) network SNR (of 16.97), and the same $\rho_{\rm{{HOM}}} / \rho_{\rm{tot}}$  (of 0.14) . Even if the posterior on the mass ratio is not significantly better, the one on the inclination is $ \sim 2 \times$ narrower. Hence, the support of the posterior in the $q$-$\iota$ plane has a smaller surface for the lensed scenario, showing that the HOMs are better constrained when we observe a lensed image pair than in the case of a single image with the same total HOM content.}
     \label{fig:HOMmeasureLensing}
 \end{figure}
 
As an important example, HOMs allow us to better constrain the orbital inclination, as seen in Fig.~\ref{fig:HOMmeasureLensing}. Hence, the detection of two lensed images with a presence of HOMs would allow us to study the HOM content with greater precision. This is likely to have implications for e.g.~the use of GW lensing in cosmology \cite{Hannuksela:2020xor}, or testing  general relativity by probing the polarization content of gravitational waves \cite{Goyal:2020bkm}.

\section{Summary and conclusions}
\label{sec:Conclusion}
In this work, we have focused on the impact lensing and HOMs can have on each other when observing a lensed image pair. We have shown that our ability to identify the strong lensing image types greatly improves when jointly analyzing two images as opposed to one. If we were to identify the presence of type-II images, it would count as smoking-gun evidence that the event is indeed lensed. In addition, we have confirmed that the presence of a type II image is required to unequivocally identify the observed image types on the basis of GW data alone. We have also shown that when the HOMs play an important role, their non-inclusion in the lensing analysis can lead to the non-detection of a lensed pair. Finally, we have shown that strongly lensed gravitational-wave events allow us to study the HOM content more accurately than similar non-lensed gravitational waves. 

\section*{Acknowledgements}
\label{sec:Acknow}
The authors thank Juan Calderon Bustillo for his very useful input about HOMs for gravitational waves and for helping to decide the example events used in this work. We are also grateful to Apratim Ganguly, Aditya Vijaykumar, and Ajit Mehta for discussion on a related topic.

JJ, OAH, and CVDB are supported by the research program of the Netherlands Organisation for Scientific Research (NWO). TGFL and ES were partially supported by grants from the Research Grants Council of the Hong Kong (Project No. CUHK 14306218), The Croucher Foundation of Hong Kong and Research Committee of the Chinese University of Hong Kong.
The authors are grateful for computational resources provided by the LIGO Laboratory and supported by the National Science Foundation Grants No. PHY-0757058 and No. PHY-0823459.

\bibliography{bibli}

\end{document}